\documentclass{article}
\usepackage{amsmath, amssymb, amsfonts}
\title{On the Vaz no horizon black hole } 
\author{Hristu Culetu, \\Ovidius University, Dept.of Physics and Electronics, \\B-dul Mamaia 124, 900527 Constanta, Romania, \\e-mail : hculetu@yahoo.com}

\begin{document}
\numberwithin{equation}{section}
\pagenumbering{arabic}
\maketitle
\newcommand{\fv}{\boldsymbol{f}}
\newcommand{\tv}{\boldsymbol{t}}
\newcommand{\gv}{\boldsymbol{g}}
\newcommand{\OV}{\boldsymbol{O}}
\newcommand{\wv}{\boldsymbol{w}}
\newcommand{\WV}{\boldsymbol{W}}
\newcommand{\NV}{\boldsymbol{N}}
\newcommand{\hv}{\boldsymbol{h}}
\newcommand{\yv}{\boldsymbol{y}}
\newcommand{\RE}{\textrm{Re}}
\newcommand{\IM}{\textrm{Im}}
\newcommand{\rot}{\textrm{rot}}
\newcommand{\dv}{\boldsymbol{d}}
\newcommand{\grad}{\textrm{grad}}
\newcommand{\Tr}{\textrm{Tr}}
\newcommand{\ua}{\uparrow}
\newcommand{\da}{\downarrow}
\newcommand{\ct}{\textrm{const}}
\newcommand{\xv}{\boldsymbol{x}}
\newcommand{\mv}{\boldsymbol{m}}
\newcommand{\rv}{\boldsymbol{r}}
\newcommand{\kv}{\boldsymbol{k}}
\newcommand{\VE}{\boldsymbol{V}}
\newcommand{\sv}{\boldsymbol{s}}
\newcommand{\RV}{\boldsymbol{R}}
\newcommand{\pv}{\boldsymbol{p}}
\newcommand{\PV}{\boldsymbol{P}}
\newcommand{\EV}{\boldsymbol{E}}
\newcommand{\DV}{\boldsymbol{D}}
\newcommand{\BV}{\boldsymbol{B}}
\newcommand{\HV}{\boldsymbol{H}}
\newcommand{\MV}{\boldsymbol{M}}
\newcommand{\be}{\begin{equation}}
\newcommand{\ee}{\end{equation}}
\newcommand{\ba}{\begin{eqnarray}}
\newcommand{\ea}{\end{eqnarray}}
\newcommand{\bq}{\begin{eqnarray*}}
\newcommand{\eq}{\end{eqnarray*}}
\newcommand{\pa}{\partial}
\newcommand{\f}{\frac}
\newcommand{\FV}{\boldsymbol{F}}
\newcommand{\ve}{\boldsymbol{v}}
\newcommand{\AV}{\boldsymbol{A}}
\newcommand{\jv}{\boldsymbol{j}}
\newcommand{\LV}{\boldsymbol{L}}
\newcommand{\SV}{\boldsymbol{S}}
\newcommand{\av}{\boldsymbol{a}}
\newcommand{\qv}{\boldsymbol{q}}
\newcommand{\QV}{\boldsymbol{Q}}
\newcommand{\ev}{\boldsymbol{e}}
\newcommand{\uv}{\boldsymbol{u}}
\newcommand{\KV}{\boldsymbol{K}}
\newcommand{\ro}{\boldsymbol{\rho}}
\newcommand{\si}{\boldsymbol{\sigma}}
\newcommand{\thv}{\boldsymbol{\theta}}
\newcommand{\bv}{\boldsymbol{b}}
\newcommand{\JV}{\boldsymbol{J}}
\newcommand{\nv}{\boldsymbol{n}}
\newcommand{\lv}{\boldsymbol{l}}
\newcommand{\om}{\boldsymbol{\omega}}
\newcommand{\Om}{\boldsymbol{\Omega}}
\newcommand{\Piv}{\boldsymbol{\Pi}}
\newcommand{\UV}{\boldsymbol{U}}
\newcommand{\iv}{\boldsymbol{i}}
\newcommand{\nuv}{\boldsymbol{\nu}}
\newcommand{\muv}{\boldsymbol{\mu}}
\newcommand{\lm}{\boldsymbol{\lambda}}
\newcommand{\Lm}{\boldsymbol{\Lambda}}
\newcommand{\opsi}{\overline{\psi}}
\renewcommand{\tan}{\textrm{tg}}
\renewcommand{\cot}{\textrm{ctg}}
\renewcommand{\sinh}{\textrm{sh}}
\renewcommand{\cosh}{\textrm{ch}}
\renewcommand{\tanh}{\textrm{th}}
\renewcommand{\coth}{\textrm{cth}}

\begin{abstract}
We propose a no horizon black hole whose collapsing matter condenses close to the event horizon and before its formation. Compared to Vaz's model \cite{CV2}, our interior geometry depends only on one parameter $r_{0}$ - the radius of the region where quantum fluctuations are significant. While the equation of state of the inner fluid is $p_{r} = -\rho$ and the traverse pressures are vanishing, the surface stress tensor corresponds to an anisotropic fluid with negative surface tension. Using the junction conditions on the boundary of the collapsing star, we found that $r_{0}$ is half its Schwarzschild radius and not $2M$, as previously obtained by Vaz for a dust ball.
 \end{abstract}
 
\section{Introduction}
Although the classical collaps process suggest that an enough massive star will undergo collapse until a singularity forms, the picture deeply changes when quantum gravity (QG) is taken into account \cite{CV1, HE, CV2, CV3}. It is supposed QG effects will play a crucial role in determining the outcome of gravitational collapse during its final stages. The semiclassical analysis would suggest that information is lost if the black hole (BH) evaporates completely through Hawking radiation. Therefore, the thermal evaporation mechanism leaves probably behind a stable remnant that contains all the information falling into the BH. Hawking \cite{SH} has recently expressed objections to the AMPS firewall \cite{AMPS, AMPSS} and suggested that the correct solution of the AMPS paradox is that event horizons do not form but only apparent horizons. Moreover, Mersini-Houghton \cite{LMH} shows that due to the negative energy Hawking radiation, the collapse of the star stops at a finite radius before the singularity and the BH horizon have formed. The star bounces instead of collapsing to a BH.

 Vaz \cite{CV2} proposed a resolution to the AMPS paradox, considering that the collapsing matter does not undergo continuous collapse to a singularity but condenses on the apparent horizon of the BH. He constructed static solutions with no tangential pressures. Every infalling shell of dust is accompanied by the emission of a positive energy shell from the center of the collapsing stellar object and this process of energy extraction from the center continues until the collapse terminates, describing the effect of strong quantum fluctuations close to the center \cite{CV2}. 

Our starting point in this paper is Vaz's quasi-classical configuration, namely a spherically-symmetric source that occupies a finite region. Instead of using a dust cloud that condenses into the apparent horizon, our fluid stress tensor possesses a radial pressure $p_{r} = -\rho$ where $\rho$ is its energy density, with no tangential pressures. The metric inside the collapsing ''plasma ball'' is simple, with a curvature singularity at $r = 0$ which, however, is not a part of the spacetime (we have taken, as Vaz \cite{CV2} did, $r \geq r_{0}$, where $r_{0}$ is the region where quantum fluctuations are expected to dominate. In addition, we introduce a surface stress tensor on the outer boundary $r_{b}$ of the star, for the junction conditions to hold. Our proposal leads to the same relation $r_{b} = 2M + r_{0}$ between the boundary radius $r_{b}$ and the Schwarzschild mass $M$ of the star but, however, we found that $r_{0} = M$ and not $r_{0} = 2M$ as the author of \cite{CV2} obtained for his dust ball.

\section{Interior metric}
Vaz \cite{CV2} already found the solutions of Einstein's equations (without $\Lambda$) corresponding to a source located in a finite region. He did not impose any equation of state between the energy density $\rho$ and the radial pressure $p_{r}$ and set the trasversal pressures to zero. However, his interior geometry depends on an extra parameter $\gamma$ and $p_{r}$ has a complicate dependence on the radial variable $r$. To make things more simple, we propose 
 \begin{equation}
p_{r} = -\rho
 \label{2.1}
 \end{equation}
as the equation of state of the fluid inside the collapsing star. It is not the case to repeat the steps used by Vaz for obtaining the expressions for $\rho$ and $p_{r}$ and the corresponding interior geometry. With the extra restriction (2.1) the following inner geometry is obtained
     \begin{equation}
   ds^{2} = -\frac{r_{0}}{r} dt^{2} + \frac{r}{r_{0}}dr^{2} + r^{2} d \Omega^{2} 
 \label{2.2}
 \end{equation}
where $d\Omega^{2}$ stands for the metric on the unit two-sphere and $r_{0}$, as we shall see, is related to the domain where the quantum fluctuations from the central region are significant. As it was noticed in \cite{CV2}, the singularity at $r = 0$ does not create problems as the solution is valid for $r \geq r_{0}$ only. 

The energy-momentum tensor looks now as
  \begin{equation}
   T^{a}_{~b} = diag\left(-\frac{1}{8\pi r^{2}}, - \frac{1}{8\pi r^{2}}, 0, 0\right)
 \label{2.3}
 \end{equation} 
where $a, b$ run from $0$ to $3$, $\rho = -p_{r} = 1/8\pi r^{2},~p_{\theta} = p{\phi} = 0$. Let us note that $T^{a}_{~b}$ from (2.3) satisfies all the energy conditions and $\rho$ and $g_{rr}$ acquire the same expressions as those obtained by Vaz because to get their values the expression of $g_{tt}$ is nowhere used. We also observe that, although the equation of state is of de Sitter type, the metric (2.2) is not de Sitter, due to the vanishing of the tangential pressures. In the domain of interest ($r \geq r_{0}$) the spacetime (2.2) has no singularities and no horizons. For example, the Ricci scalar is $R^{a}_{~a} = 2/r^{2}$ and the Kretschmann scalar is $R_{abcd}R^{abcd} = 4(r^{2} - 2r_{0}r + 3r_{0}^{2})/r^{6}$, with a maximum value of $8/r_{0}^{4}$ at $r = r_{0}$. Moreover, the radial acceleration of a static observer is $a^{r} = -r_{0}/2r^{2} < 0$, so that the gravitational field is repulsive.  

The total mass $m(r)$ by the radius $r$ is given by \cite{SW}
  \begin{equation}
	m(r) = \int{4\pi r^{2}\rho(r)dr} = \frac{r}{2}
 \label{2.4}
 \end{equation} 
as if inside any sphere of radius $r$ were a BH with mass $m(r)$. As far as the Komar energy is concerned, we obtain \cite{TP, HC}
  \begin{equation}
	  W = 2 \int(T_{ab} - \frac{1}{2} g_{ab}T^{c}_{~c})u^{a} u^{b} N\sqrt{det\gamma} d^{3}x = 0,
 \label{2.5}
 \end{equation} 
due to the contribution of the negative radial pressure. $N = \sqrt{-g_{tt}}$ is the lapse function, $u^{a} = (\sqrt{r/r_{0}}, 0, 0, 0)$ is the velocity vector field of a static observer and $det\gamma$ represents the determinant of the spatial metric. 

The Misner-Sharp mass $M_{ms}$ is obtained from \cite{CG, NY, HPFT}
  \begin{equation}
  1 - \frac{2M_{ms}}{r} = g^{ab} \nabla_{a}r ~\nabla_{b}r,
 \label{2.6}
 \end{equation}
which gives us
  \begin{equation}
  M_{ms} = \frac{r}{2} - \frac{r_{0}}{2}.
 \label{2.7}
 \end{equation}
If we denote $r_{b}$ the radius of the boundary of the star, we have $M_{ms}(r_{b}) = r_{b}/2 - r_{0}/2$ which, as we shall see, gives exactly the Schwarzschild mass $M$ measured by an outer observer. In other words, the total mass \cite{SW} $m(r) = M_{0} + M_{ms}$, so that $M_{0} = r_{0}/2$ gives the (negative) contribution of the central mass \cite{CV2}.
 
 \section{Junction conditions}
 As we mentioned above, $r_{b}$ gives the outer boundary of the collapsing star. Our next step is to match the interior spacetime (2.2) to the exterior empty space described by the Schwarzschild geometry
 \begin{equation}
   ds^{2} = -(1 - \frac{2M}{R}) dT^{2} + \frac{1}{1 -\frac{2M}{R}}dR^{2} + R^{2} d \Omega^{2}  
 \label{3.1}
 \end{equation}
 where $M$ is the Schwarzschild mass, measured from large distances. Equating the 1st fundamental forms of (2.2) and (3.1), one obtains 
  \begin{equation}
  \frac{r_{0}}{r_{b}}dt^{2} = (1 - \frac{2M}{R_{b}}) dT^{2},~~~\frac{r_{b}}{r_{0}}dr^{2} = \frac{1}{1 -\frac{2M}{R_{b}}}dR^{2},~~~r_{b} = R_{b}
 \label{3.2}
 \end{equation}
 whence
   \begin{equation}
   \sqrt{\frac{r_{0}}{r_{b}}}t_{b} = \sqrt{1 -\frac{2M}{R_{b}}}T_{b},~~~\sqrt{\frac{r_{b}}{r_{0}}(1 - \frac{2M}{r_{0}})} = 1.
 \label{3.3}
 \end{equation}
  Using the 2nd relation (3.3) we conclude that $T_{b} = t_{b}$ and therefore we have
  \begin{equation}
   r_{b} = r_{0} + 2M
 \label{3.4}
 \end{equation} 
i.e. the same condition as that infered by Vaz \cite{CV2} (his Eq. 24), but using a different interior geometry. Eq. (3.4) shows that the star radius $r_{b}$ is always greater than the Schwarzschild radius, even at the final stages of the collapsing process. The difference is exactly the radius of the region of strong quantum fluctuations.

For an estimation of $r_{0}$ we call on the 2nd junction condition - the relation between the jump of the extrinsic curvature when the boundary is crossed and the surface stress tensor (the Lanczos equation)
     \begin{equation}
     [K_{ab}] - h_{ab}[K^{c}_{~c}] = -8\pi S_{ab}
 \label{3.5}
 \end{equation} 
where $[K_{ab}] = K^{+}_{ab} - K^{-}_{ab}$ is the jump of the extrinsic curvature of the $r = r_{b}$ boundary $\Sigma$, $h_{ab} = g_{ab} - n_{a}n_{b}$ is the induced metric on $\Sigma$ and $n_{b}$ is the normal on $\Sigma$, with $n_{b}n^{b} = 1$. The extrinsic curvature tensor is given by
  \begin{equation}
 K^{\pm}_{ab} = h^{c}_{~a}\nabla_{c}n^{\pm}_{b},    
 \label{3.6}
 \end{equation} 
where $+(-)$ refers to the exterior (interior) geometries. With $n^{-}_{a} = (0, \sqrt{r/r_{0}}, 0, 0)$, the inner geometry gives us
  \begin{equation}
  K^{-}_{tt} = \frac{r_{0}}{2r^{2}}\sqrt{\frac{r_{0}}{r}},~~~K^{-}_{\theta \theta} = \sqrt{r_{0}r},~~~K^{-,a}_{a} = \frac{3}{2r}\sqrt{\frac{r_{0}}{r}},
 \label{3.7}
 \end{equation} 
evaluated at $r = r_{b}$. With $n^{+}_{a} = (0, 1/\sqrt{1 - \frac{2M}{r}}, 0, 0)$, the outer Schwarzschild geometry yields
  \begin{equation}
  K^{+}_{TT} = -\frac{M}{R^{2}}\sqrt{1 - \frac{2M}{R}},~~~K^{+}_{\theta \theta} = R \sqrt{1 - \frac{2M}{R}},~~~K^{+,a}_{a} = \frac{2M - 3R}{R^{2}\sqrt{1 - \frac{2M}{R}}}, 
 \label{3.8}
 \end{equation} 
evaluated at $r = r_{b}$. The jump of the extrinsic curvature when $\Sigma$ is crossed yields
  \begin{equation}
  [K^{a}_{a}] = \frac{1}{2r_{b}}\sqrt{\frac{r_{b}}{r_{0}}}.
 \label{3.9}
 \end{equation} 
In addition, we have
  \begin{equation}
  [K_{tt}] = -\frac{1}{2r_{b}}\sqrt{\frac{r_{0}}{r_{b}}},~~~[K_{\theta \theta}] = 0.
 \label{3.10}
 \end{equation} 
 
 The next step is to propose a stress tensor on the boundary $\Sigma$. Let $S_{ab}$ be given by the expression
   \begin{equation}
 S_{ab} = (p_{s} + \sigma )u_{a}u_{b} + p_{s}h_{ab} + \pi_{ab},  
 \label{3.11}
 \end{equation} 
 where $p_{s} = -\tau$ is the surface pressure, $\sigma$ is the surface energy density, $\tau$ is the surface tension and $ \pi_{ab}$ is the anisotropic stress tensor, with $\pi^{a}_{~a} = 0,~\pi^{a}_{~b}n^{b} = 0$. We consider the surface fluid has to be anisotropic to match the interior of the ''plasma ball'', where $p_{\theta} = 0$. By means of the equations (3.5), (3.9) and (3.10), one obtains
 \begin{equation}
 S_{tt} = 0,~~~8\pi S_{\theta \theta} =  \frac{r_{b}}{2}\sqrt{\frac{r_{b}}{r_{0}}}. 
 \label{3.12}
 \end{equation}  
 Let us assume now that
  \begin{equation}
  p_{s} + \sigma = 0
 \label{3.13}
 \end{equation}  
is the equation of state of the surface fluid. It is justified by the repulsive character of the inner gravitational field which led to $a^{r} < 0$, i.e. to keep a test particle at rest we must act on it with a force toward the center. Using now (3.12) and (3.13) in (3.11), we get the set of equations
  \begin{equation}
  8\pi (r_{b}^{2}p_{s} + \pi_{\theta \theta}) = \frac{r_{b}}{2}\sqrt{\frac{r_{b}}{r_{0}}},~~~\pi_{tt} - \frac{r_{0}}{r_{b}}p_{s} = 0,~~~\pi^{t}_{t} + 2\pi^{\theta}_{ \theta} = 0,
 \label{3.14}
 \end{equation}  
 which leads to
  \begin{equation}
  p_{s} = -\sigma = - \pi^{t}_{t} = 2\pi^{\theta}_{ \theta} = \frac{1}{24\pi r_{b}}\sqrt{\frac{r_{b}}{r_{0}}}
 \label{3.15}
 \end{equation}   
 Let us make use now of the Young-Laplace equation \cite{CDR, HC}
   \begin{equation}
   -[p_{r}] = \tau K^{a}_{~a}.
 \label{3.16}
 \end{equation}  
 With $[p_{r}] = p_{r,out} - p_{r,in} = 1/8\pi r^{2}$ from (2.3), $\tau = \sigma$ from (3.15) and $K^{a}_{~a}$ from (3.8), we obtain
   \begin{equation}
 \sqrt{\frac{r_{b}}{r_{0}}} (2r_{b} - 3M) = 3r_{b}\sqrt{\frac{r_{0}}{r_{b}}}  
 \label{3.17}
 \end{equation}  
 whence
   \begin{equation}
   2r_{b} - 3M = 3r_{0}.
 \label{3.18}
 \end{equation}   
 Solving for $r_{0}$ in (3.4) and (3.18), we finally come to
    \begin{equation}
  r_{0} = M.
 \label{3.19}
 \end{equation}   
 In other words, $r_{0}$ is half the gravitational radius of the Schwarzschild mass $M$. In terms of the mass $M$ the surface energy density and surface pressure become
   \begin{equation}
  p_{s} = -\sigma  = \frac{1}{24\pi \sqrt{3}M}.
 \label{3.20}
 \end{equation}   
 
 For the Misner-Sharp mass we get $M_{ms} = r_{b}/2 - r_{0}/2 = M$ and the total mass $m(r_{b}) = r_{b}/2 = 3M/2$. To summarize, the negative central mass $M_{0} = r_{0}/2$ is the difference between the total mass $m(r_{b})$ till the boundary and the Misner-Sharp mass. We note that our result does not coincide with $r_{0} = r_{b}/2$ obtained by Vaz \cite{CV2}. His estimation of $r_{0}$ is based on how much energy is extracted from the center during the star collapse (every collapsing shell is accompanied by an inner, outgoing wave which will extract energy from the center). However, in the calculation of the average energy of the outgoing shell, Vaz used a power series developing which seems, in our opinion, to be debatable. For example, the r.h.s. of Eq. (27) of \cite{CV2} is obtained in the limit the shell spacing $\sigma \rightarrow 0$. More precisely, the small parameter is considered to be $\sigma r_{i}$ so that, actually,  $\sigma r_{i} << 1$ (private communication from the author of \cite{CV2}). But we must take in fact  $\sigma r_{i}/l^{2}_{Pl} << 1$, for to use a dimensionless expansion parameter ($l_{Pl}$ is here the Planck length). But it is hard to conceive a realistic shell to obey such an inequality, with $l_{Pl}^{2}$ at the denominator. Therefore, we consider the region of strong quantum fluctuations occupies half the Schwarzschild radius of the collapsing star.
 
 \section{Conclusions}
 Recently Hawking \cite{SH} raised several objections to the firewall formation during the collapse of a black hole. The final stage of the collapsing process is not the central singularity but only an apparent horizon forms. Based on Hawking conjecture, Vaz \cite{CV2, CV3} considers dust collapse that terminates on the apparent horizon. In his model the collapse wave function  indicates that there is a process by which energy extraction from the center takes place.

We proposed in this paper a similar mechanism for a no horizon black hole formation by means of a simple interior geometry where the equation of state of the fluid is $p_{r} + \rho = 0$. To satisfy the junction conditions on the boundary of the ''plasma ball'', we introduced a surface stress tensor corresponding to an anisotropic fluid. Using the Young-Laplace equation as an extra condition, we found that the radius $r_{0}$ of the region where there are strong quantum fluctuations is half the gravitational radius of the Schwarzschild BH, contrary to Vaz's result that $r_{0} = 2M$.


\begin{thebibliography} {14}

\bibitem{CV1}
C. Vaz, Gen. Relat. Grav. 43, 3429 (2011) (arXiv: 1111.6821 [gr-qc]).
\bibitem{HE}
S. W. Hawking and G. F. R. Ellis, \textit{The large scale structure of spacetime}, Cambridge University Press, Cambridge (1973).
\bibitem{LMH}
L. Mersini-Houghton, arXiv: 1406.1525 [hep-th].
\bibitem {CV2}
C. Vaz, arXiv: 1407.3823 [gr-qc].
\bibitem{CV3}
C. Vaz, Int. J. Mod. Phys. D23, 1441002 (2014) (arXiv: 1405.4898 [gr-qc]).
\bibitem{SH}
S. W. Hawking, arXiv: 1401.5761 [hep-th].
\bibitem{AMPS}
A. Almheiri, D. Marolf, J. Polchinski and J. Sully, JHEP 1302, 062 (2013) (arXiv: 1207.3123 [hep-th]). 
\bibitem{AMPSS}
A. Almheiri, D. Marolf, J. Polchinski, D. Stanford and J. Sully, arXiv: 1304.6483 [hep-th]). 
\bibitem{SW}
S. Weinberg, \textit{Gravitation and Cosmology}, Wiley, 1972, p. 302.
\bibitem{TP}
T. Padmanabhan, Class. Quantum Grav. 21, 4485 (2004) (arXiv: gr-qc/0308070).
\bibitem{HC}
H. Culetu, Int. J. Mod. Phys. Conf. Ser. 3, 455 (2011) (arXiv: 1304.5386 [gr-qc]).
\bibitem{CG}
M. Carrera and D. Giulini, Rev. Mod. Phys. 82, 169 (2010) (ArXiv: 0810.2712 [gr-qc]).
\bibitem{NY}
A. Nielsen and D.-h. Yeom, Int. J. Mod. Phys. A24: 5261 (2009) (ArXiv: 0804.4435 [gr-qc]).
\bibitem{HPFT}
 L. Herrera et al., Int. J. Mod. Phys. D18: 129 (2009) (ArXiv: 0804.3584 [gr-qc]).
\bibitem {CDR}
V. Cardoso, O. Dias and J. Rocha, JHEP 1001, 021 (2010) (arXiv: 0910.0020 [hep-th]).



\end{thebibliography}
\end{document}